\documentstyle[psfig]{aanew}

\begin{document}


   \title{Mkn 463 field observed by \emph{Beppo}SAX}

\author{R.~Landi\inst{1,2} and L.~Bassani\inst{1}}

\offprints{R.~Landi (landi@tesre.bo.cnr.it)}

\institute{ITeSRE/CNR, via Piero Gobetti 101, I--40129 Bologna, Italy
\and 
Dipartimento di Fisica, Universit\`a di Bologna, via Irnerio
46, I--40129 Bologna, Italy}
\date{Received / accepted}
\abstract{
In this work we present the observation of the Mkn 463
field performed with the MECS instrument on--board \emph{Beppo}SAX in the
1.8--10.5 keV band. The Mkn 463 field is an example of an
extragalactic field crowded with absorbed X--ray sources: 
apart from the Seyfert 2 galaxy Mkn 463 and the well known QSO PG 1352+183 (the only 
object showing no absorption), two
other objects are detected with a column density in excess to the galactic value. 
The first 1SAX J1353.9+1820 is a red QSO from
the \emph{Beppo}SAX High Energy Large Area Survey (HELLAS). The second
1SAX J1355.4+1815 is optically unidentified, but its X--ray spectral
characteristics indicate that it too is an AGN hidden behind a large
column density.  
\keywords{X--rays: galaxies --
                Galaxies: Seyfert --
                Galaxies: individual: Mkn 463
               }}
\authorrunning{R.~Landi $\&$ L.~Bassani}
\titlerunning{Mkn 463 field observed by \emph{Beppo}SAX}
\maketitle 
%
%
%
%
%
%
%
\section{Introduction}
Obscured active galactic nuclei (AGN) or type 2 objects, in which the
power source lies behind a significant column density,
are increasingly found in X--ray surveys as instruments become more and
more sensitive.
How common these objects are is still quite uncertain. 
Optical and Far--Infrared based estimates
suggest a ratio of Seyfert 2 to Seyfert 1 ranging from 2 to 4 
(Osterbrock $\&$ Martel 1993; Ho, Filippenko $\&$ Sargent 1997;
Schmitt 2001), but these values very likely 
underestimate the number of heavily absorbed objects (those with 
$N_{\rm H}>10^{24}$ cm$^{-2}$, or Compton thick) which
can be identified only with X--ray surveys; in particular there may be as
many Compton thick Seyfert 2s as Compton thin ones 
(Bassani et al. 1999; Risaliti et al. 1999), 
implying a Seyfert 2 to Seyfert 1 ratio in the range 3--5. This
fraction can be even larger based on a simple argument recently 
discussed by Matt et al. (2000). The three nearest AGN (CenA, NGC 4945 and
the Circinus Galaxy all within 4 Mpc) are all heavily obscured with 
column densities $\ge$~10$^{23}$ cm$^{-2}$. The probability of
finding a relatively unobscured AGN so close is only 0.05, 
i.e. 60 times less than observed.
Only if obscured AGN outnumber unobscured ones by about a factor of 10 
to one does the above probability exceed 2$\%$.
Here we show observationally how common absorbed sources can be. During a
\emph{Beppo}SAX
observation of the sky region surrounding Mkn 463, a highly absorbed
Seyfert 2 galaxy itself,
we have found two other nearby sources which are characterized by a high 
column density.
A fourth object detected in the same field was found to be of type 1.
The probability of finding an absorbed X--ray source in a MECS field of
view of 0.8 square degrees can be 
estimated on the basis of the \emph{Beppo}SAX HELLAS (\emph{High Energy 
Large Area Survey}) survey Log $N$--Log $S$ (Vignali 2000).
This survey detects a higher fraction ($\sim$ 20$\%$) of absorbed AGN
with respect to other X--ray surveys.
At the 2--10 keV flux limit of $\sim$~(1--2)~$\times10^{-13}$
erg cm$^{-2}$ s$^{-1}$ of our observation, we expect one HELLAS source per
field of view (or 0.2 absorbed objects), i.e. at least a factor of 3 lower 
than observed.
Again our results point to a large fraction of absorbed objects in 
the local universe.
The true value of this fraction will eventually emerge as the sky is surveyed 
at progressively better sensitivity and/or higher X--ray frequencies.
\section{Observation and data reduction}
This work concerns the observation of the Seyfert 2 Mkn 463
field performed by the the Medium Energy Concentrator Spectrometer (MECS;
Boella et al. 1997a) on--board \emph{Beppo}SAX (Boella et al. 1997b). 

The observation was divided into three separate shots of the same
field of view:
the effective exposure times for the three observations are
$5.3\times10^{3}$ s (Obs. 1, July 3rd 1998), $28.4\times10^{3}$ s
(Obs. 2, July 5th 1998) and $53.2\times10^{3}$ s (Obs. 3, 28th 1998)
respectively.

The target source, Mkn 463, was intentionally not on--axis, 
so as to also include
in the MECS field of view a nearby quasar bright in
X--rays, PG 1352+183. 

The two longest observations (Obs. 2 and Obs. 3) resulted in a very
complex and interesting field of view; the third 
one was too short for any convincing evidence of sources.
In order to improve the imaging quality of the data, 
the two longest observations were summed to produce a combined
image of this sky region.  

In the 1.8--10.5 keV band (Fig. 1) two more sources were detected 
by the MECS in addition to Mkn 463 and PG 1352+183. 

With respect to previous observations, the most interesting result is the 
detection of a new X--ray source, 1SAX J1355.4+1815.  
The source is
located at $\alpha(2000)=13^{h}55^{m}26.4^{s}$,
$\delta(2000)=+18^{\circ}16'32.0^{\prime \prime}$,
$\sim$~10$'$
south of Mkn 463. The uncertainty associated with the
source position is 1$'$ (90$\%$ confidence level).
This source is not identified optically neither in NED nor SIMBAD, 
although it appears as 1AXG 135526+1816 in the \emph{ASCA}--GIS 
catalogue (Ueda et al. 2001) with a 2--10 keV 
flux of $2.7\times10^{-13}$ erg cm$^{-2}$ s$^{-1}$. 
At soft X--ray energies, the object is listed as 1WGA J1355.4+1816
with a count rate ranging from 0.002--0.006, i.e. a factor of 50 or so
dimmer than PG 1352+183. 
Finally, a fourth source located at 
$\alpha(2000)=13^{h}53^{m}55.8^{s}$, 
$\delta(2000)=+18^{\circ}19'36.8^{\prime \prime}$
is identified with a previously
discovered \emph{Beppo}SAX object belonging to the HELLAS survey:
1SAX J1353.9+1820 optically identified with a broad line AGN at
$z=0.217$.
The H$\alpha$ equivalent width ($\sim$ 90 \AA), the absence of H$\beta$
and the \emph{Beppo}SAX hardness ratio indicate significant reddening of 
the nuclear radiation. Furthermore, the presence of a broad Balmer line 
and the properties of the optical spectrum allow to classify this object 
as a fairly low--luminosity red quasar (Vignali et al. 2000).
In the following, we describe the spectral properties of all four sources
detected during this observation.
\begin{table*}[t]
 \centering
 \caption[]{\emph{Beppo}SAX MECS sources count rates and normalization factors
of the two longest observations (Obs. 2 and Obs. 3) in the Mkn 463 field.}
 \begin{tabular}{l|c|c|c}
\hline
   &Obs. 1     &Obs. 2 & Normalization Factor\\
   & Count rate ($\times10^{-3}$ s$^{-1}$)&Count rate ($\times10^{-3}$ s$^{-1}$)&  \\
\hline
     &    &    &     \\
Mkn 463  & $1.61\pm0.41$   &$2.4\pm0.3$ &  0.59$^{+0.39}_{-0.31}$  \\  
    &    &   &     \\
PG 1352+183 & $7.68\pm0.77$ &$6.0\pm0.6$   & 1.20$^{+0.30}_{-0.26}$ \\
   &     &    &    \\
1SAX J1355.4+1815 &$4.85\pm0.65$ & $5.2\pm0.5$ &0.86$^{+0.26}_{-0.23}$\\
    &   &    &    \\
1SAX J1353.9+1820&$1.02\pm0.34$ & $0.9\pm0.2$& 0.90$^{+0.94}_{-0.62}$\\
    &    &   &     \\
\hline
\end{tabular}
\end{table*}
\section{Spectral analysis}
For each individual observations, spectral data were extracted from 
regions centered on each source whose radius depends on the source
intensity: we choose a radius of 4$'$ for 
1SAX J1355.4+1815 and PG 1352+183 and 2$'$ for
Mkn 463 and 1SAX J1353.9+1820.
%
%
%
%
\begin{figure}
\psfig{file=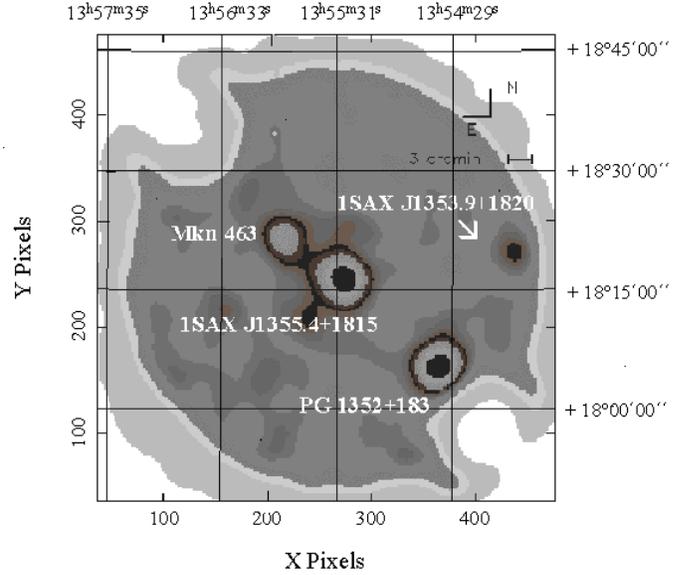,width=9.2cm,height=8cm,angle=0}
\caption{MECS image of the sky region surrounding Mkn 463 obtained
by summing Obs. 2 and Obs. 3.}
\end{figure}

The background subtraction for the on--axis source was performed
using
blank sky spectra extracted from the same region of the source.
Because of their faintness, for the off--axis sources we used
both "blank sky" background spectra and local background spectra
(extracted from a region with a radius analogous to the source extraction
radius), finding no significant difference between the background
subtracted spectra obtained with the two methods. Thus, we decided to use
in our spectral analysis the local background for the off--axis sources.

The data were rebinned in order to sample the energy
resolution of
the detector, with an accuracy proportional to the count rate. The
spectral analysis has been performed by means of the {\sc XSPEC 10.0}
package, and using the instrument response matrices released by the
\emph{Beppo}SAX Science Data Center in September 1997. 
For the off--axis sources we used the appropriate ancillary response
files to avoid the effects of the vignetting due to the mirrors. 
All quoted errors
correspond to 90$\%$ confidence intervals for one interesting parameter
($\Delta\chi^2=2.71$).
Unfortunately individual observations provided poor quality data to allow
a spectral study of each source. For this reason, we
tested the possibility of summing observations together also for spectral
analysis.
However, since variability on long timescales is common in active galaxies
and the measurements were taken in quite different periods, we first
analyzed,
for each source and for each instrument, the two 
longest data sets together (Obs. 2~+~Obs. 3),
but allowing for a normalization constant in the simple power law fit 
adopted in the first instance. The results of this test listed in
Table 1 indicate that the normalization factors are always compatible with  
1 within errors and therefore we combined Obs. 2 and Obs. 3 also for spectral
analysis.

The absorption of X--rays due to our galaxy in the
direction of each object, which amounts 
to $\sim$ $2\times10^{20}$ cm$^{-2}$ (Dickey \& Lockman, 1990), is fixed
in all models used for the spectral analysis.
\subsection{1SAX J1355.4+1815: an unidentified X-ray source}
We first fit the data
with a single power law to search for any extra feature: this fit
gives a slightly flat index ($\Gamma=1.45^{+0.32}_{-0.29}$) and a
$\chi^{2}/\nu=11.2/17$.
If we introduce
intrinsic absorption in the source the fit is improved
($\chi^{2}/\nu=9.8/16$,
see Fig. 2), but the spectrum is more canonical being characterized by a
$\Gamma=1.95^{+0.10}_{-0.70}$ and by an absorbing column
density $N_{\rm H}=\big(1.73^{+2.90}_{-1.73}\big)\times10^{22}$
cm$^{-2}$.
Inspection of Fig. 2 indicates
that some residual emission may be present around 6--7 keV, suggesting
the introduction in the fit of a narrow gaussian emission line. The
line turns out to be centered at $6.5\pm0.3$ keV and has an
equivalent width ($EW$) of 1.4$^{\rm+0.7}_{\rm-1.0}$ keV.
Fixing the line rest frame
energy at 6.4 keV provides an upper limit to the
source redshift of 0.03.
The addition of the line provides an improvement in the fit
($\Delta\chi^{2}=5.4$ for two additional parameters) which is 
significant at more than the 90$\%$ confidence level. 
If the line width is allowed to vary the best fit value is
0.24$^{+0.52}_{-0.24}$ keV, i.e. consistent with being zero.  
The fit with the line gives a too steep power law ($\Gamma=2.5$) for an
AGN, albeit the uncertainty is large; 
if the power law index is fixed to 1.9, a value more
appropriate to an AGN X--ray spectrum, the column density reduces to
$(1.1\pm1.0)\times10^{22}$ cm$^{-2}$ and the line $EW$ becomes
920$^{+650}_{-714}$ eV ($\chi^{2}/\nu=7.4/15$).
The 2--10 keV observed flux is $4.6\times10^{-13}$ erg
cm$^{-2}$ s$^{-1}$, corresponding to an absorption--corrected luminosity
$\le 2\times10^{42}$ erg s$^{-1}$. The observed flux suggests
a variation of a factor of $\sim$~2 with respect to the \emph{ASCA}/GIS 
data.
The line $EW$ is too high ($\ge$~200 eV) to be produced by transmission
in the
observed absorbing medium, indicating the presence of either a higher
column density than observed or a strong reflection component. 
Following Turner et al. (1997), the observed $EW$ 
can be produced by transmission and/or reflection in an absorbing cold
medium with a 
column density of the order of $\sim$ 10$^{23-24}$ cm$^{-2}$.
Therefore    
1SAX J1355.4+1815 could be another example of an AGN hidden behind
large amount of gas and dust.
\subsection{Mkn 463}
Mkn 463 was the original target of the observation: this Seyfert 2 galaxy
has been observed by \emph{ASCA}, discussed by Ueno et al. (1996) and
more recently reanalyzed by
Levenson et al. (2001). The \emph{ASCA} spectrum was well described
by a three components model (the thermal plus scattering components 
which reproduce the soft
X--ray spectrum and an absorbed power law which provides 
the high energy emission)
yielding a 2--10 keV flux of $6.6\times10^{-13}$ erg
cm$^{-2}$ s$^{-1}$. The measured column density 
was about $3\times10^{23}$ cm$^{-2}$.
An iron line was not evident in the data with an upper limit to the
$EW$ of 890 eV.
Fitting the \emph{Beppo}SAX data with a simple absorbed power law provides 
an unrealistic inverted
spectral index of --~0.35 and no absorption ($\chi^{2}/\nu=19.5/16$). 
Fixing the spectral index to a more realistic
value of 1.9 requires extra absorption with respect to the galactic value
of 
$\big(3.6^{+2.1}_{-1.4}\big)\times10^{23}$ cm$^{-2}$
($\chi^{2}/\nu=24.3/17$).
The fit is however not good implying a wrong modeling of the source continuum.
In view of the \emph{ASCA} results, we introduced in the 
fit a scattering component 
with photon index equal to the one of the primary emission, i.e. 1.9.
In this case, we obtain an absorption of
$\big(6.0^{+4.1}_{-3.8}\big)\times10^{23}$ cm$^{-2}$.
An iron line is required/called for with a high statistical 
significance (96$\%$, $\chi^{2}/\nu=12.8/14$):
the line energy is 6.58$^{+0.23}_{-0.48}$ keV with an 
$EW=1.15^{+0.77}_{-0.94}$ keV,
indicative of a heavily absorbed AGN.
The 2--10 keV flux was $1.1\times10^{-13}$ erg cm$^{-2}$ s$^{-1}$, 
a factor of $\sim$ 2 higher than the previous \emph{ASCA} 
observation. Again the line properties as well as the ratio of the X--ray
flux compared to 
the [O{\small III}] and Far--Infrared flux point to a heavily absorbed AGN
(Bassani et al. 1999), i.e. a source with a column density 
$\sim$~10$^{24}$ cm$^{-2}$.
\begin{figure}
\psfig{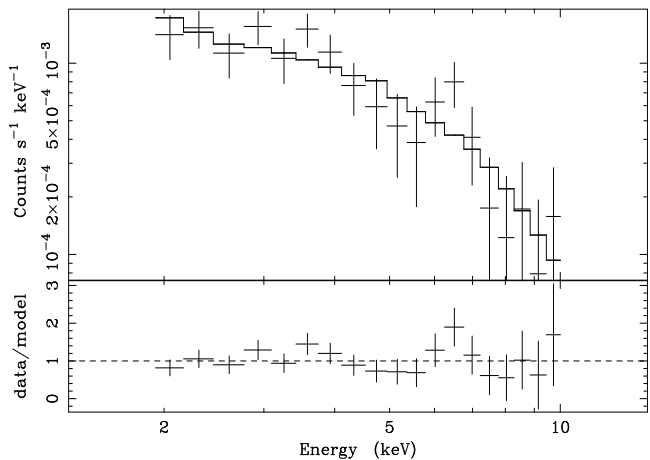}
\caption{Data to model ratio when MECS data of 1SAX J1355.4+1815
are fitted with a simple power law.}
\end{figure}
\subsection{PG 1352+183}
PG 1352+183 is a low redshift ($z=0.158$) quasar previously observed
by \emph{Beppo}SAX on January 1998 (Mineo et al. 2001). The source
spectrum is well represented 
by a  power law  with $\Gamma=2.3$ and a 2--10 keV flux of 
$1\times10^{-12}$ erg
cm$^{-2}$ s$^{-1}$. A cold iron line was also detected at a 
confidence level of 97$\%$; the line $EW$ was estimated to be in the
range 
300--1200 eV. The source is detected in each of our observations
and is in fact 
the strongest object in the field of view. We have performed a spectral
analysis on the combined data similar to that of Mineo and co--workers, 
and we obtain very similar results except for the  
higher flux observed: ($\Gamma=2.3^{+1.1}_{-0.9}$,  
$N_{\rm H}\le 5.2\times10^{22}$ cm$^{-2}$ and a 2--10 keV flux 
of $1.8\times10^{-12}$ erg cm$^{-2}$ s$^{-1}$). 
The line is also detected with a  
significance of $\sim$~92~$\%$; if the line energy is fixed at 6.4
keV, 
the line equivalent width is 680$^{+770}_{-680}$ eV.  
The source obviously changed in flux by a factor of $\sim$~2 over a
timescale of
six months which is quite common in active galaxies, without any evidence
of strong variation in the spectral shape.
The line equivalent width is higher (albeit the uncertainty is large)
than that usually observed in Seyfert 1 
galaxies, but similar to what found in other PG quasars (Mineo et al.
2001).
\subsection{1SAX J1353.9+1820}
This source has also been observed by \emph{ASCA} twice  and found to be 
characterized by an absorbed power law ($\Gamma$~=~1.2--1.3) with a
column density of 
the order of (4--6)~$\times10^{21}$ cm$^{-2}$. The flux remained stable
between the
two observations at a level of (4--6)~$\times10^{-13}$ erg cm$^{-2}$
s$^{-1}$. 
Despite the low statistic available, we fitted the data 
in order to check consistency with previous observations.
We first fit the data with a simple power law model (see Fig. 3).
The fit is satisfactory 
($\chi^{2}/\nu=16.9/17$) and gives a very flat spectrum characterized
by a photon index $\Gamma$ $\le$ 0.87 which is in agreement with the  
\emph{ASCA} results (Vignali et al. 2000). 
Fixing the photon index to 1.9, the absorbing
column density turned out to be of 
$\big(4.7^{+7.0}_{-3.5}\big)\times10^{22}$ cm$^{-2}$, only
marginally compatible with that measured by \emph{ASCA}.

The unabsorbed 2--10 keV flux and luminosity are $\sim$ 
$3.6\times10^{-13}$ erg cm$^{-2}$ s$^{-1}$ and $4.6\times10^{43}$
erg s$^{-1}$. The flux value is lower by a factor of two with respect to
that measured with \emph{ASCA}, while the                 
5--10 keV flux is $2.4\times10^{-13}$
erg cm$^{-2}$ s$^{-1}$, i.e. $\sim$~30$\%$ lower than the value found
in the previous \emph{Beppo}SAX observation (Fiore et al. 1999).\\
\begin{figure}
\psfig{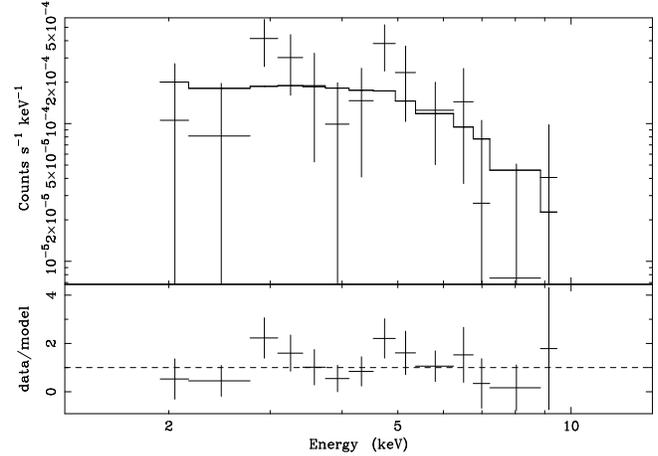}
\caption{Data to model ratio when MECS data of 1SAX J1353.9+1820 are
fitted
with a simple power law plus galactic absorption.}
\end{figure}
\section{Conclusions}
The sky region surrounding Mkn 463 was found to be populated with X--ray
emitting objects: apart from Mkn 463 itself (the original purpose of the 
observation) and a nearby well known QSO, we also detected two more 
sources. 
The first of the two is optically unidentified and
is likely to be an absorbed object on the basis of the spectral shape and 
emission line characteristics, the second one is optically classified as a 
red QSO mildly absorbed in X--rays.
The present study provides observational evidence for the high fraction 
of absorbed objects expected to be present among the AGN population.
This result is based on simple qualitative arguments as well as on
estimates of the AGN contribution 
to the cosmic X--ray background. Overall, the ratio of type 2 to type 1
objects
is expected to exceed 3--5; this is the ratio we observed in our
observation.
Our objects are, however, local and still of low luminosity in contrast 
to the more powerful version of the Seyfert 2 galaxy population 
or the long sought type 2 QSOs. The situation 
is now changing with Chandra and XMM which have both begun to uncover such 
powerful obscured AGN (Hasinger et al. 2001; Norman et al. 2001;
Tozzi et al. 2001). 
\begin{acknowledgements}
This research has made use of SAXDAS linearized and cleaned event 
files produced at
the \emph{Beppo}SAX Science Data Center. 
Financial support from the Italian Space Agency is gratefully 
acknowledged. We would like to thank the referee Dr. A. Orr for the very
useful comments which have improved the quality of this work. 
\end{acknowledgements}

\begin{thebibliography}{}
   \bibitem[1999]{Bassa} Bassani, L., Dadina, M., Maiolino, R., et al.
1999, ApJS, 121, 473
   \bibitem[1997]{Boella97a} Boella, G., Butler, R.~C., Perola, G.~C., et
al. 1997a, A\&AS, 122, 299
   \bibitem[1997]{Boella97b} Boella, G., Chiappetti, L., Conti, G., et al.
1997b, A\&AS, 122, 327
   \bibitem[1990]{Dick90} Dickey, J.~M., Lockman, F.~J. 1990, ARA\&A, 28,
215 
   \bibitem[1999]{Fio99} Fiore, F., La Franca, F., Giommi, P., et al.
1999, MNRAS, 306, L55 
   \bibitem[1997]{Fro97} Frontera, F., Costa, E., dal Fiume, D., et al.
1997, A\&AS, 122, 357
   \bibitem[2001]{Hasing} Hasinger, G., Altieri, B., Arnaud, M., et al.
2001, A\&AS, 365, L45
   \bibitem[1997]{HoFil} Ho, L.~C., Filippenko, A.~V., $\&$ Sargent,
W.~L.~W. 1997, ApJS, 112, 315
   \bibitem[2001]{Lev} Levenson, N.~A., Weaver, K.~A., Heckman, T.~M.
2001, ApJS, 133, 269
   \bibitem[2000]{Matt1} Matt, G., Fabian, A.~C., Guainazzi, M., et al.
2000, MNRAS, 318, 173
   \bibitem[2001]{Mineo} Mineo, T., Fiore, F., Laor, A., et al. 2000,
A\&A, 359, 471
   \bibitem[2001]{Norman} Norman, C., Hasinger, G., et al. 2001,
submitted to ApJ~[astro--ph/0103198]
   \bibitem[1993]{Osterb} Osterbrock, D.~E., $\&$ Martel, A. 1993, ApJ,
414, 552  
   \bibitem[1997]{Parm97} Parmar, A.~N., Martin, D.~D.~E., Bavdaz, M., et
al. 1997, A\&AS, 122, 309
   \bibitem[1999]{Risali} Risaliti, G., Bassani, L., Comastri, A., et al.
1999, Mem. Soc. Astron. Ital., 70, 73
   \bibitem[2001]{Schmit} Schmitt, H.~R. 2001, BAAS, 198, 3601
   \bibitem[2001]{Tozzi1} Tozzi, P., Rosati, P., et al. 2001, ApJ in 
press [astro--ph/0103014]
   \bibitem[1997]{T97} Turner, T.~J., George, I.~M., Nandra, K., et al.
1997, ApJ, 418, 164
   \bibitem[2001]{Ueda01} Ueda, Y., Ishisaki, Y., Takahashi, T., et al.
2001, ApJS, 133, 1
   \bibitem[1996]{Ueno1} Ueno, S., Koyama, K., Awaki, H., Hayashi, I., \& 
Blanco, P.~R. 1996, PASJ, 48, 389 
   \bibitem[1998]{Vig98} Vignali, C., Comastri, A., Stirpe, G.~M., et al.
1998, A\&A, 333, 411
   \bibitem[2000]{Vigna00} Vignali, C., Mignoli, M., Comastri, A., 
Maiolino, R., \& Fiore, F. 2000, MNRAS, 314, 11
   \bibitem[2000]{Vig2000} Vignali, C. 2000, PhD thesis, University of
Bologna
\end{thebibliography}
\end{document}